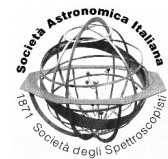

# GRB Cosmography with THESEUS

L. Izzo[1], M. Muccino[2], M. Della Valle[3,2]

[1] Instituto de Astrofisica de Andalucia, Glorieta de la Astronomia s/n, 18008, Granada, Spain, e-mail: `izzo@iaa.es`
[2] ICRANet, P.zza della Repubblica 10, 65122 Pescara, Italy,
[3] INAF-Osservatorio Astronomico di Napoli, Salita Moiariello, 16, I-80131 Napoli, Italy

**Abstract.** Gamma-ray Bursts can be used as distance indicators using the Combo relation. We show how the proposed THESEUS mission will allow us to investigate the evolution history of the Universe at high redshift with GRBs. Assuming that THESEUS will measure the redshift for 800 GRBs, we show that the accuracy on the cosmological parameters of the main cosmological models will greatly improve, so that we can use GRBs as additional and independent cosmological probes and also put strong constraints on the evolution of the dark energy equation of state parameter $w(z)$.

## 1. Introduction

Gamma-ray Bursts (GRBs) are the most luminous explosions in the Universe (Gehrels et al. 2009), a characteristic that allows them to be observed at very high redshifts (Tanvir et al. 2009). GRBs are classified according to their $T_{90}$ duration (the time duration where the 90% of the total GRB emission is observed) in "long" and "short" (Kouveliotou et al. 1993): the first class contains GRBs whose $T_{90}$ is larger than 2 s, in the second one there are GRBs with $T_{90} \leq 2$ s. This division is also physical: long GRBs likely originate from the collapse of very massive stars (Woosley & Heger 2006), they are associated with broadline Ic supernovae (SNe) (Cano et al. 2017), they have a softer prompt emission and are mainly observed in star-forming dwarf galaxies (Fruchter et al. 2006; Blanchard et al. 2016). Short GRBs are instead associated to the merging of two compact objects (Berger 2014), an hypothesis confirmed by the recent detection of a gravitational wave event and a kilonova event coincident with the short GRB 1708017A (Abbott et al. 2017; Smartt et al. 2017; Tanvir et al. 2017).

All these characteristics render GRBs very powerful probes for the high-redshift cosmology. In particular, they are among the only probes of the star-formation rate (Kistler et al. 2009; Trenti et al. 2012) and of the physical properties (metallicity, abundances, ionization) of the first galaxies observed at high redshifts. In addition to these characteristics, GRBs can be powerful tracers of the evolution history of the Universe, but this task implies the use of them as distance indicators.

In recent years, the existence of some correlations among GRB observables has been definitely confirmed: the total energy emitted in the GRB prompt emission correlates with the energy where the high-energy prompt emission peaks (Amati et al. 2002) and these two quantities have been found to correlate together with the total isotropic X-ray emission in the GRB afterglow (Bernardini et al. 2012). Recently, we have proposed a new method that takes origin from these two correlations and where the use of the afterglow plateau luminosity is proposed as possible "candle" for determining GRB distances (Izzo et al. 2015). The plateau luminosity $L_0$ correlates with the rest-frame peak energy of the prompt emission $E_{p,i}$ and with the characteristic transition time $\tau$ in the X-ray afterglow between the plateau and the late power law (decay index $\alpha_X$). The analytical formulation is the following:

$$log\left(\frac{L_0}{erg/s}\right) = log\left(\frac{A}{erg/s}\right) + \quad (1)$$
$$+ \gamma \left[log\left(\frac{E_{p,i}}{keV}\right) - \frac{1}{\gamma}log\left(\frac{\tau/s}{|1+\alpha_X|}\right)\right]$$

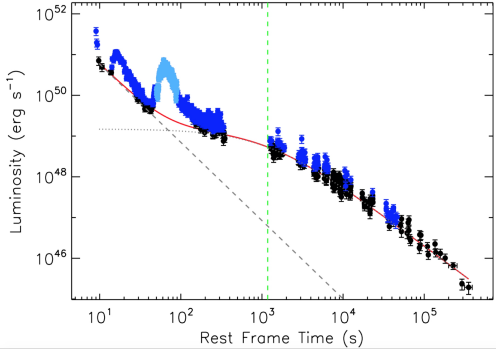

**Fig. 1.** The fit of the Swift XRT luminosity light curve of the afterglow of GRB 140206A. To determine the plateau flux $F_0$ a similar fit is done on the rest-frame (0.3 - 10) keV flux of the GRB. The dashed green line corresponds to the value of $\tau$.

We named this correlation "Combo", being the combination between the Amati and the Bernardini relations. In Izzo et al. (2015) we have compiled a first sample which included 60 GRBs up to December 2014. The GRB distance is then inferred from the plateau luminosity $L_0$ estimated from Eq. 1 and the corresponding observed flux $F_0$, determined through a light-curve fitting procedure of the X-ray afterglow. This is modelled with an iterative procedure that recomputes the best-fit after excluding possible flaring activity until the best-fit reaches a p-value larger than 0.3 (Izzo et al. 2015). In Fig. 1 it is shown as an example the results of the fit of the X-ray luminosity light curve of GRB 140206A.

The initial sample of 60 GRBs has been expanded recently (Muccino et al. in preparation), and now it is composed by 132 GRBs up to December 2016. The most interesting feature is the absence of outliers in the correlation, but not all long GRBs are included in our sample, due to the absence of a measured $E_{p,i}$ or an incomplete X-ray afterglow light curve for these events. In Fig. 2 it is shown the Combo relation computed assuming a $\Lambda$ Cold Dark Matter ($\Lambda$CDM) cosmological model for the computation of all GRB distances, and consequently for their luminosities $L_0$.

## 2. Cosmological applications

One of the most important applications of the Combo relation consists in studying the evolution history of the Universe through estimate of the cosmological parameters. As note above, it is possible to use $L_0$ as distance indicator once we estimate from the observed data the plateau flux $F_0$. Using the updated sample of 132 GRBs, which ranges a redshift interval from $z = 0.145$ to $z = 8.5$, we have tested different cosmological models with GRBs but here we present the results for only two cases: 1) the $\Lambda$CDM case with a non-zero value for the curvature parameter ($\Omega_k \neq 0$) and 2) the case of a redshift-evolving Dark Energy equation of state (EOS) $w$, in the contest of a flat scenario ($\Omega_m = 1 - \Omega_\Lambda$).

For the $\Lambda$CDM case we obtain results in complete agreement with other probes like supernovae Ia (Riess et al. 1998; Perlmutter et al. 1999): after marginalizing over the Hubble constant value, which we fix at the value of $H_0 = 73.03 \pm 1.79$ (Riess et al. 2016), we obtain for the matter density parameter $\Omega_m = 0.28^{+0.09}_{-0.07}$. However, if we "thaw" the Dark Energy EOS $w$ parameter from the standard $w = -1$ value, under the flatness condition of the Universe ($\Omega_{tot} = \Omega_m + \Omega_\Lambda = 1$), we obtain interesting results: while the standard $w = -1$ value is still satisfied up to redshift $z \sim 3$, at higher distances this condition is valid only at 2-$\sigma$ of confidence, see also Fig. 3. This result implies the necessity of additional analysis using the Combo dataset, and the extension to a larger number of GRBs will greatly improve our accuracy on the cosmological parameters.

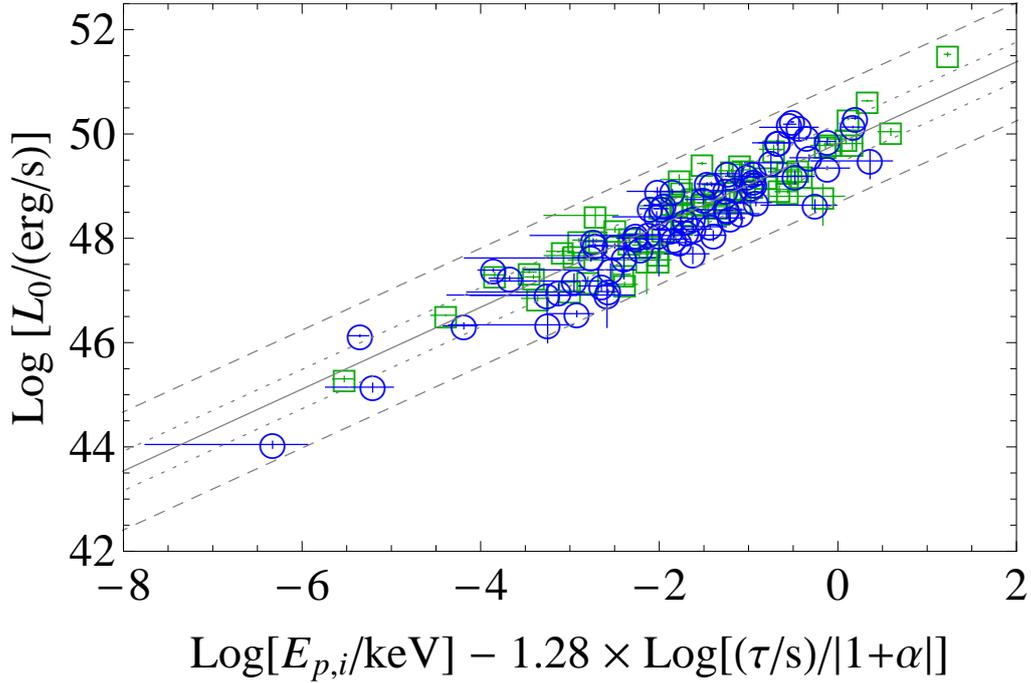

**Fig. 2.** The Combo relation in the ΛCDM model from the 60 GRBs considered in Izzo et al. (2015) (green squares) and the new 72 ones (blue circles) that will be presented soon (Muccino et al. in preparation). The solid black best-fit line and the 1- and 3- extra scatter $\sigma_{ex}$ of the relation (dotted and dashed gray lines) are also displayed.

## 3. Simulations with THESEUS

The Transient High Energy Sky and Early Universe Surveyor (THESEUS, Amati et al. (2017)) is a proposed mission for the investigation of the high-energy Universe and among its main goals there is the detection and the measurement of the redshift for a large number of GRBs, of the order of 800 events, in particular GRBs at very large distances. Given also the possibility of determining on-board the redshift of GRBs, thanks to the presence of a near-infrared telescope on the payload, and the additional presence of a Soft X-ray Imager operating in the (0.3 - 6) keV energy range, THESEUS represents the ideal instrument to investigate the history evolution of the Universe with the Combo relation.

With a sample of 800 expected GRBs with a redshift known, we can largely improve the accuracy on the cosmological parameters and on the Dark Energy EOS. To demonstrate the previous claim, we made a MonteCarlo simulated sample of 800 GRBs whose observed properties fall inside the 3-$\sigma$ confidence limit for each of the parameters of the Combo relation, as computed for the updated sample of 132 GRBs, with the errors on these parameters fixed to be the $\sim$ 20% of the simulated values. We then have built a corresponding Hubble Diagram for this simulated sample, see fig. 4, which we used for the estimate of the cosmological parameters. We have applied this method for the case of the $\Lambda$CDM model as well as for a flat Universe with a constant Dark Energy equation of state, usually called also *wCDM* model. Our results are very well accurate, see Fig. 5, and show how GRBs can actually used to constrain cosmological parameters with a very good precision: for the first model we infer $\Omega_m = 0.26 \pm 0.03$ while for the second one $\Omega_m = 0.24 \pm 0.08$ and $w = -0.92 \pm 0.30$. This latter parameter is fully consistent with the $\Lambda$CDM cosmological constant value $w = -1$.

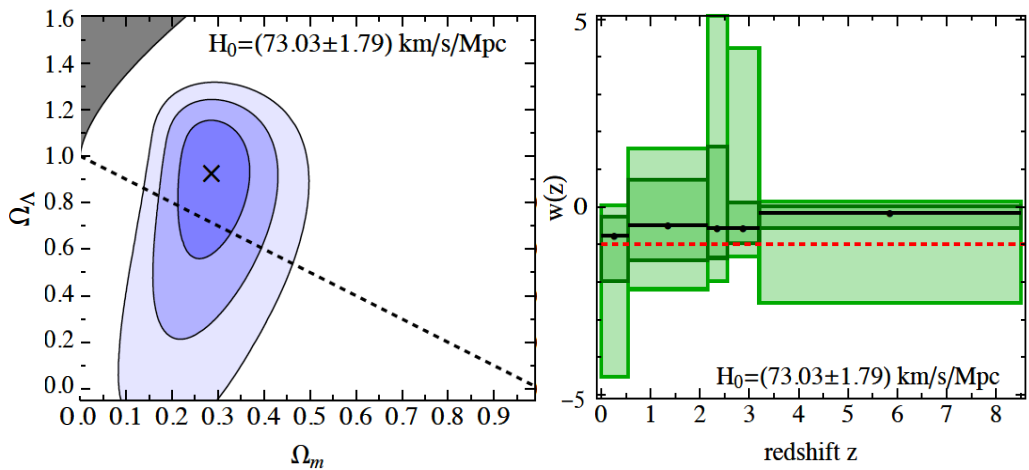

**Fig. 3.** (Left panel) The best-fit result (black cross) and the confidence regions (1-, 2-, and 3- from the inner/darker to the outer/lighter) in the $\Omega_m - \Omega_\Lambda$ plane for the $\Lambda$CDM cosmological model. The gray regions indicate the set of parameters for which no Big Bang occurs and the dashed lines those for a flat Universe. (Right panel) The evolution of the Dark Energy EOS parameter $w$ as a function of the redshift, in the case of a flat Universe with $\Omega_m = 0.43^{+0.09}_{-0.11}$.

## 4. Conclusions

With the advent of a new era of detectors for high-energy astronomy, we will improve more and more our knowledge on the phenomenon of GRBs. The detailed analysis of GRBs in the last 20 years with new techniques and instruments has led to the formulation of precise correlations between GRB observables. One of the consequences of this achievement consists in the possibility of using GRBs as distance indicators. The recently proposed Combo re-lation is revealing as one of the most promising tool for investigating the evolution history of the Universe (Izzo et al. 2015). In the near future, the THESEUS mission will largely increase the number of observed GRBs at all redshifts, allowing us to investigate with an unprecedented accuracy the possible evolution of the Dark Energy EOS at all redshifts. This will establish GRBs as an independent distance indicators useful for testing all cosmological models, and as additional check for the recent tension on the Hubble constant value $H_0$, recently reported from CMB observations (Planck Collaboration et al. 2016) and supernovae Ia (Riess et al. 2016).

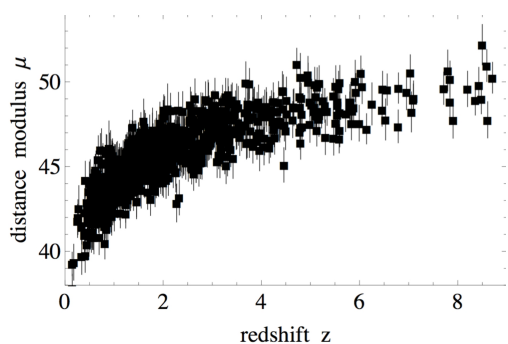

**Fig. 4.** The Hubble Diagram obtained from the MC simulated sample of 800 GRBs with known redshift, as expected from the THESEUS mission after five years, using the Combo relation as distance idicator method for the estimate of the GRB distances.

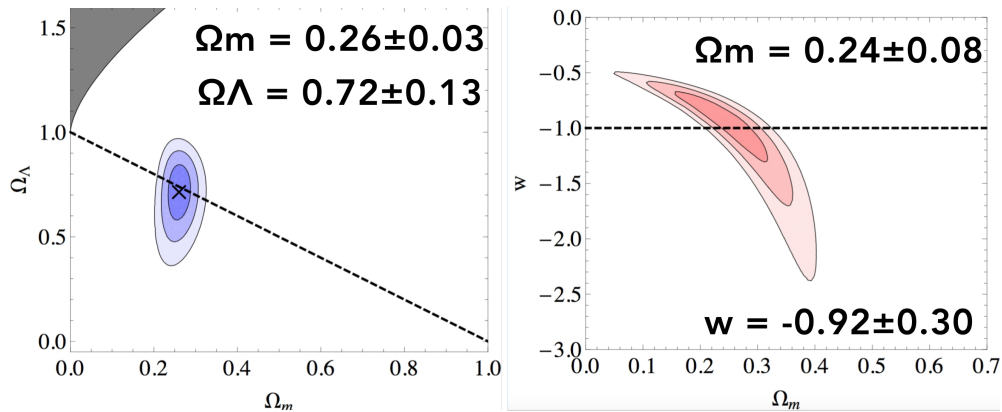

**Fig. 5.** (Left panel) The best-fit result (black cross) and the confidence regions (1-, 2-, and 3- from the inner/darker to the outer/lighter) in the $\Omega_m - \Omega_\Lambda$ plane for the $\Lambda$CDM cosmological model and considering the MC-simulated sample of 800 GRBs with known redshift, as expected from the THESEUS mission after five years. (Right panel) The best-fit result (black cross) and the confidence regions (1-, 2-, and 3- from the inner/darker to the outer/lighter) in the $\Omega_m - w$ plane for the wCDM cosmological model.

*Acknowledgements.* LI7 acknowledges support from the Spanish research project AYA 2014-58381-P. MM acknowledges the partial support of the Ministry of Education and Science of the Republic of Kazakhstan.